\documentclass[ reprint,nofootinbib,amsmath,amssymb,aps]{revtex4-1}
\usepackage{graphicx}
\usepackage{rotating}
\usepackage{dcolumn}
\usepackage{bm}
\usepackage{color}
\usepackage{mathptmx, textcomp}
\usepackage[latin1]{inputenc}
\usepackage{braket}
\usepackage[colorlinks,linkcolor=magenta,anchorcolor=cyan,citecolor=blue]{hyperref}
\usepackage[multiple]{footmisc}

\def\be{\begin{equation}}
\def\ee{\end{equation}}
\def\ba{\begin{eqnarray}}
\def\ea{\end{eqnarray}}

\newcommand{\eq}[1]{(\ref{#1})}

\def\q{\theta} \def\r {\rho}     \def\p {\pi} \def\a {\alpha}  \def\d {\delta}  \def\g {\gamma} \def\h {\eta}   \def\l {\lambda}    \def\b {\beta}   \def\pd {\partial}\def\p {\pi} \def \inf {\infty}  
\def\Q{\Theta}       \def\D {\Delta}  \def\G {\Gamma} \def\L {\Lambda}    \def\grad{\nabla}\def\.{\cdot}
\def\math {\mathcal}
\begin{document}

\title{{Generalized Covariant Entropy Bound in Lanczos-Lovelock Gravity}}
\author{Ming Zhang$^1$}
\email{mingzhang@jxnu.edu.cn}
\author{Jie Jiang$^2$}
\email{Corresponding author. jiejiang@bnu.edu.cn}
\affiliation{${}^1$Department of Physics, Jiangxi Normal University, Nanchang 330022, China}
\affiliation{${}^2$College of Education for the Future, Beijing Normal University, Zhuhai 519087, China}
\date{\today}

\begin{abstract}
In this paper, we investigate the generalized covariant entropy bound in the theory where the Einstein gravity is perturbed by the higher-order Lovelock terms. After replacing the Bekenstein-Hawking entropy with the Jacobson-Myers entropy and introducing two reasonable physical assumptions, we showed that the corresponding generalized covariant entropy bound is satisfied under a higher-order approximation of the perturbation from the higher-order Lovelock terms. Our result implies that the Jacobson-Myers entropy strictly obeys the entropy bound under the perturbation level, and the generalized second law of Lanczos-Lovelock gravity is also satisfied when the Einstein gravity is perturbed by the higher-order Lovelock terms.

\end{abstract}
\maketitle
\section{Introduction}
The investigation of black hole thermodynamics has led to some interesting entropy bounds that should be observed to guarantee theoretical consistency. Bekenstein \cite{Bekenstein:1980jp} has conjectured that the entropy $S$ and energy $E$ of any stable gravitational thermodynamic system satisfies a universal bound
\ba\begin{aligned}
S\leq 2\p E R\,,
\end{aligned}\ea
in which $R$ is defined as the circumferential radius of the sphere surrounding the thermodynamical system. This bound is called the Bekenstein bound and it can be indicated by the generalized second law (GSL) of black holes. The Bekenstein bound has been confirmed in many weakly gravitational systems with finite size. In a strongly gravitational system, it is hard to define the energy $E$ and radius $R$ locally. Counterexamples can be found in the process of gravitational collapse \cite{Bousso:1999xy}. For a spherical system in Einstein gravity, the Bekenstein bound can be simplified as
\ba\begin{aligned}
S\leq\frac{A}{4}\,,
\end{aligned}\ea
in which $S$ and $A$ are the entropy and area of the system. It is worth noting that this bound is not well defined in a strongly gravitational system since the area $A$ is dependent on the choice of the spacelike region in the system and it can always be selected to be arbitrarily small by an almost null hypersurface. It was shown that this bound can be violated in the system for large volume \cite{Fischler:1998st}.

To find a covariant version of the entropy bound, Bousso considered the entropy cross a light sheet and proposed a covariant entropy bound, called the Bousso bound \cite{Bousso:1999xy}, which can be well formulated in arbitrarily curved spacetime. Consider a $(D-2)$-dimensional compact spacelike surface $B$ with area $A(B)$. Let $L$ be a null hypersurface generated by the null geodesics which starts at $B$ and is orthogonal to $B$. Assume that the expansion of the null congruence is nonpositive (i.e., $L$ is a light sheet) and $L$ is not terminated until a caustic point is reached. Then, the entropy $S_L$ passing through the light sheet $L$ is bounded by the quarter of $A(B)$, i.e.,
\ba\begin{aligned}
S_L\leq \frac{A(B)}{4}\,.
\end{aligned}\ea
This is the covariant bound proposed by Bousso and it is conjectured to be valid in any strongly gravitational system with arbitrary large regions. This bound is shown to be hold in various cases \cite{Flanagan:1999jp,Gao:2004mc,Gao:2005bk,Gao:2008zy,Bousso:2003kb,Pesci:2007rp,Pesci:2008yy} and it can be regarded as a formulation of holographic principle in spacetime.

Note that the above conjecture requires that the light sheet $L$ ends at a caustic point. Flanagan $et.\,al.$ \cite{Flanagan:1999jp} extended this bound in which the light sheet can be terminated at another $(D-2)$-dimensional spatial surface $B'$ before reaching a caustic. Then  the entropy bound is modified as
\ba\begin{aligned}\label{ebound}
S_L\leq \frac{1}{4}\left|A(B')-A(B)\right|\,,
\end{aligned}\ea
in which $A(B')$ is the area of the spatial surface $B'$. This is called the generalized covariant entropy bound or generalized Bousso bound and it has been proved in Einstein gravity under some physical assumptions \cite{Flanagan:1999jp,Bousso:2003kb,Strominger:2003br}.

General relativity (GR) is not a complete theory of gravity due to the lack of a definitive quantum gravity theory and it can only be regarded as an effective theory in a certain region of scale. After considering the quantum effect or string modification, the higher-curvature terms are often added to Einstein-Hilbert action to modify the effective action of the gravitational theory \cite{A13,A14,A15,A16}. In these cases, the Einstein gravity is perturbed by the higher-curvature terms. In this paper, we focus on the Lanczos-Lovelock gravity, which is the only natural generalization of Einstein gravity to higher-dimensional spacetime if we demand that the equations of motion are second-order differential equations of metric \cite{Lanczos,Lovelock}. Moreover, unlike most higher-curvature gravitational theories, Lanczos-Lovelock gravity is a ghost-free theory and admits consistent initial value formulation \cite{Lovelock2,Kovacs:2020ywu}. As mentioned above, the generalized covariant entropy bound \eq{ebound} is only valid in Einstein gravity. It is natural to ask whether the higher-curvature corrections can affect the entropy bound of the gravitational theory. Recently, Matsuda $et.\, al.$ \cite{Matsuda:2020yvl} extended the generalized covariant entropy bound into the modified gravitational theory by replacing the quarter of area $A(B)/4$ with some appropriate black hole entropy $S_\text{bh}(B)$, such as the Wald entropy \cite{A17,A18} or Jacobson-Myers (JM) entropy \cite{Jacobson:1993xs}. Under two reasonable assumptions, they proved the entropy bound for Wald entropy in $f(R)$ gravity and canonical scalar-tensor theory. Moreover, they also showed that the bound using JM entropy holds for the GR branch of spherically symmetric configurations in Einstein-Gauss-Bonnet gravity. In the following, we would like to extend their discussion into the case where the Einstein gravity is perturbed by the higher-order Lovelock terms and show that the JM entropy can give a reasonable entropy bound in this theory.

The outline of this paper is as follows. In Sec. \ref{sec2}, we briefly review the Lanczos-Lovelock gravity and discuss the features of Wald entropy and JM entropy. In Sec. \ref{sec3}, we introduce the generalized entropy bound in Lanczos-Lovelock gravity and show the physical assumptions as well as the key point for proving this bound. In Sec. \ref{sec4}, we prove the generalized entropy bound in the theory where the Einstein gravity is perturbed by the higher-order Lovelock terms and show that the JM entropy strictly obeys the entropy bound under the perturbation level. Finally, the conclusion and discussion are presented in Sec. \ref{sec5}.

\section{Lanczos-Lovelock gravity}\label{sec2}

In this paper, we consider the Lanczos-Lovelock gravitational theory with some minimally coupled matter fields. The action of this theory in $D$-dimensional spacetime is given by
\ba\begin{aligned}
I=\frac{1}{16\p}\int d^Dx\sqrt{g}\left(\sum_{k=0}^{k_\text{max}}\frac{a_k}{2^k}\math{L}^{(k)}+\math{L}_\text{mat}\right),
\end{aligned}\ea
in which $\math{L}_\text{mat}$ is the Lagrangian density of the matter fields, $g_{ab}$ is the Minkowski metric of the spacetime, and
\ba\begin{aligned}
\math{L}^{(k)}=\d_{c_1d_1\cdots c_kd_k}^{a_1b_1\cdots a_k b_k}R^{c_1d_1}_{a_1b_1}\cdots R^{c_kd_k}_{a_kb_k}
\end{aligned}\ea
is the $k$-order Lovelock term. Here $k_\text{max}=\left[(D-1)/2\right]$\footnote{The Square brackets [x] denotes the integer part of x.} and
\ba\begin{aligned}
\d_{c_1d_1\cdots c_kd_k}^{a_1b_1\cdots a_k b_k}=(2k)!\d^{[a_1}_{c_1}\d^{b_1}_{d_1}\cdots\d^{a_k}_{c_k}\d^{b_k]}_{d_k}
\end{aligned}\ea
is the generalized Kronecker tensor. The equation of motion is given by
\ba\begin{aligned}
&E_{ab}=8\p T_{ab}\,,
\end{aligned}\ea
in which $T_{ab}$ is the stress-energy tensor  of the matter fields, and
\ba\begin{aligned}
E_{ab}=-\sum^{k_\text{max}}_{k=0}\frac{a_k}{2^{k+1}}\d_{ac_1d_1\cdots c_kd_k}^{b a_1b_1\cdots a_k b_k}R^{c_1d_1}_{a_1b_1}\cdots R^{c_kd_k}_{a_kb_k}
\end{aligned}\ea
is the generalized Einstein tensor of Lanczos-Lovelock gravity. Employing the Noether charge method of Iyer and Wald \cite{A17,A18}, the Wald entropy of Lanczos-Lovelock gravity can be obtained and it is given by
\ba\begin{aligned}
S_W=-2\p\int_s d^{D-2}x\sqrt{\g}P^{abcd}\hat{\bm{\epsilon}}_{ab}\hat{\bm{\epsilon}}_{cd}\,,
\end{aligned}\ea
where we have denoted
\ba\begin{aligned}
P_{ab}^{cd}=\frac{1}{16\p}\sum^{k_\text{max}}_{k=0}\frac{k a_k}{2^k}\d_{cdc_2d_2\cdots c_kd_k}^{aba_2b_2\cdots a_k b_k}R^{c_2d_2}_{a_2b_2}\cdots R^{c_kd_k}_{a_kb_k}\,.
\end{aligned}\ea
Here $s$ is a cross-section of event horizon, $\g_{ab}$ is the induced metric on $s$, and $\hat{\bm{\epsilon}}_{ab}$ is the binormal to $s$. The Wald entropy gives the correct first law in the stationary black holes. However, as discussed in Refs. \cite{A25,A26,A27,A28}, the Wald entropy of the Lanczos-Lovelock gravity does not obey the linearized second law and we need to focus on the JM entropy, i.e.,
\ba\label{JMS}
S_\text{JM}=\frac{1}{4}\int_s d^{D-2}x\sqrt{\g} \r_\text{JM}
\ea
with
\ba\begin{aligned}
\r_\text{JM}=\sum^{k_\text{max}}_{k=1}\frac{k a_k}{2^{k-1}}\d_{c_2d_2\cdots c_kd_k}^{a_2b_2\cdots a_k b_k}\hat{R}^{c_2d_2}_{a_2b_2}\cdots \hat{R}^{c_kd_k}_{a_kb_k}\,.
\end{aligned}\ea
in which $\hat{R}^{cd}_{ab}$ is the Riemann tensor of the induced metric $\g_{ab}$ on the cross-section $s$. In the stationary black hole, JM entropy and Wald entropy give the same result, and therefore the JM entropy also obeys the first law. Considering the relationship between the generalized covariant entropy bound and the generalized second law of the black holes, it is natural to apply the JM entropy to discuss the entropy bound in the Lanczos-Lovelock gravity.

\section{Generalized covariant entropy bound}\label{sec3}

In this section, we first introduce the basic setups of the generalized covariant entropy bound in Lanczos-Lovelock gravity. Let $L$ be a null hypersurface generalized by null geodesics, which starts at a compact $(D-2)$-dimensional spatial surface $B_0$ and ends at another compact $(D-2)$-dimensional spatial surface $B_1$. Let $k^a=(\pd/\pd u)^a$ be the tangent vector field of the null geodesics, in which $u$ is an affine parameter of the null geodesics such that the spatial surfaces $B_0$ and $B_1$ are given by $u=0$ and $u=1$, separately. Any spatial surface $B$ determined by the same $u$ is called the cross-section of the null hypersurface. Then, we can choose $(u, x)$ to a coordinate system on the null hypersurface $L$, in which $x=\{x^1,\cdots x^{D-2}\}$ denotes the coordinate of the cross-section and every geodesic is determined by a constant $x$. Then, the covariant entropy bound in Lanczos-Lovelock gravity demands that the entropy $S_L$ passing through the null hypersurface $L$ should satisfy
\ba\begin{aligned}\label{eb}
S_L\leq |S_\text{JM}(B_0)-S_\text{JM}(B_1)|\,,
\end{aligned}\ea
in which $S_\text{JM}(B)$ is evaluated by the JM entropy formula \eq{JMS} on the cross-section $B$.

To prove the entropy bound, we first define the generalized expansion $\Q$ of the JM entropy as the change of entropy per unit area, i.e.,
\ba\begin{aligned}
\frac{d S_\text{JM}}{d u}=\frac{1}{4}\int_B d^{D-2}x \sqrt{\g} \Q\,.
\end{aligned}\ea
Noting that the JM entropy is a purely spatial quantity in the $(D-2)$-dimensional slice $B$, i.e., it is determined by the induced metric $\g_{ab}$, we can regard $S_\text{JM}$ as an action on the cross-section $B$. Then, the Lie-derivative $\math{L}_k=\pd_u$ can be seen as a variation on $S_\text{JM}$. After assuming that $B$ is compact and dropping the surface terms in $\pd_u S_\text{JM}$, we can get
\ba\begin{aligned}
\pd_u S_\text{JM}&=-\frac{1}{4}\sum^{k_\text{max}}_{k=1}k a_k \int_{B}d^{D-2}x\sqrt{\g}[\hat{E}^{(k-1)}]^{ab}\pd_u\g_{ab}\\
&=-\frac{1}{2}\sum^{k_\text{max}}_{k=1}k a_k \int_{B}d^{D-2}x\sqrt{\g}[\hat{E}^{(k-1)}]^{ab}K_{ab}
\end{aligned}\ea
in which
\ba\begin{aligned}
[\hat{E}^{(k)}]^b_a=-\frac{1}{2^{k+1}}\d_{ac_1d_1\cdots c_kd_k}^{b a_1b_1\cdots a_k b_k}\hat{R}^{c_1d_1}_{a_1b_1}\cdots \hat{R}^{c_kd_k}_{a_kb_k}\,,
\end{aligned}\ea
and
\ba
K_{ab}=\frac{1}{2}\pd_u \g_{ab}
\ea
is the extrinsic curvature associated with $k^a$. These results imply that
\ba\begin{aligned}\label{exptheta}
\Q=K_b^a\sum^{k_\text{max}}_{k=1}\frac{k a_k}{2^{k-1}}\d_{ac_2d_2\cdots c_kd_k}^{b a_2b_2\cdots a_k b_k}\hat{R}^{c_2d_2}_{a_2b_2}\cdots \hat{R}^{c_kd_k}_{a_kb_k}
\end{aligned}\ea
after neglecting the total-derivative terms.

Choose $u$ to be an affine parameter of the null geodesics. Using the equation of motion, we can write the change of $\Q$ as
\ba\begin{aligned}\label{dQdu}
\frac{d\Q}{d u}=-8\p \math{T}+\math{F}\,,
\end{aligned}\ea
in which
\ba\begin{aligned}
\math{T}&=T_{ab}k^a k^b\,,\\
\math{F}&=H_{ab}k^a k^b+k^a \grad_a \Q\,.
\end{aligned}\ea
This can be regarded as the Raychaudhuri equation in Lanczos-Lovelock gravity.

In the thermodynamic limit, there exists an entropy flux vector field $s^a$ such that the entropy passing through the null hypersurface $L$ can be written as
\ba\begin{aligned}
S_L=\int_{L}d^{D-2}x du\sqrt{\g} s
\end{aligned}\ea
with the entropy density
\ba
s=-k_a s^a\,.
\ea

Analogies to the assumptions in Einstein gravity \cite{Bousso:2003kb,Strominger:2003br}, Ref. \cite{Matsuda:2020yvl} made two following assumptions in the modified gravitational theories,
\ba\begin{aligned}\label{assumption}
&\text{(i)}\quad \pd_u s(x,u)\leq 2\p \math{T}(x,u)\,,\\
&\text{(ii)}\quad s(x, 0)\leq -\frac{1}{4}\Q(x, 0)\,
\end{aligned}\ea
on the null hypersurface $L$. The first assumption is from the requirement that the change rate of the entropy flux is not large than the energy flux and it can also be regarded as the consequence of the version of Bekenstein bound \cite{Matsuda:2020yvl}. The second assumption is just an initial choice of the hypersurface such that the entropy bound is valid at the beginning of $L$.

With the above setups and assumptions, it is not hard to get
\ba\begin{aligned}
s(x, u)&=s(x, 0)+\int_{0}^{u} du\pd_u s(x, u)\\
&\leq s(x, 0)+2\p\int_{0}^{u}du \math{T}(x, u)\,,
\end{aligned}\ea
in which we have used the first assumption at the last step. Then, using Eq. \eq{dQdu} and together with the second assumption in Eq. \eq{assumption}, we have
\ba\begin{aligned}
s(x, u)\leq&s(x, 0)-\frac{1}{4}\Q(x, \l)+\frac{1}{4}\Q(x, 0)+\frac{1}{4}\int_{0}^{u}d\tilde{u} \math{F}(x, \tilde{u})\\
\leq&-\frac{1}{4}\Q(x, \l)+\frac{1}{4}\int_{0}^{u}d\tilde{u} \math{F}(x, \tilde{u})\,.\\
\end{aligned}\ea
Finally, after integrating the above identity over $L$, we have
\ba\begin{aligned}\label{ineq11}
S_L\leq S_\text{JM}(B_0)-S_\text{JM}(B_1)+\frac{1}{4}\int_{0}^{1}du \int d^{D-2}x\sqrt{\g}\D(x, u)\,,\quad
\end{aligned}\ea
in which we have denoted
\ba\begin{aligned}
\D(x, u)=\int_{0}^{u}d\tilde{u} \math{F}(x, \tilde{u})\,.
\end{aligned}\ea
From the above results, we can see that the key point to examining the generalized covariant entropy bound is to judge the sign of $\math{F}(x, u)$. If we have $\math{F}(x, u)\leq 0$ on $L$, the inequality \eq{ineq11} reduces to
\ba\begin{aligned}
S_L\leq S_\text{JM}(B_0)-S_\text{JM}(B_1)\,,
\end{aligned}\ea
which is the entropy bound given by Eq. \eq{eb}. For the Einstein gravity, Eq. \eq{dQdu} is just the Raychaudhuri equation and we have $\math{F}\leq 0$, which gives the proof of the generalized covariant entropy bound in Einstein gravity. In the following, we would like to judge the sign of $\math{F}$ in the Lanczos-Lovelock gravity.

\section{Proof of the entropy bound with higher-curvature corrections}\label{sec4}

From the perspective of quantum corrections and string theory, it is natural to consider the models of gravity where the Einstein gravity is perturbed by higher curvature terms. Therefore, in the following, we consider the Lanczos-Lovelock gravity where the higher-order Lovelock terms are treated as small corrections to the Einstein gravity, i.e., we consider the Lovelock theory with $a_0=-2\L$, $a_1=1$ and $a_k =\l \a_k$ for $k\geq 2$, in which $\l$ is a small quantity which describes the perturbation from the higher-curvature terms. Then, we have
\ba\begin{aligned}
H_a^b=G_a^b-\L \d_a^b-\l\sum^{k_\text{max}}_{k=2}\frac{\a_k}{2^{k+1}}\d_{ac_1d_1\cdots c_kd_k}^{b a_1b_1\cdots a_k b_k}R^{c_1d_1}_{a_1b_1}\cdots R^{c_kd_k}_{a_kb_k}\,,\quad
\end{aligned}\ea
and
\ba\begin{aligned}
\r_\text{JM}=1+\l \r
\end{aligned}\ea
with
\ba
\r=\sum^{k_\text{max}}_{k=2}\frac{k \a_k}{2^{k-1}}\d_{c_2d_2\cdots c_kd_k}^{a_2b_2\cdots a_k b_k}\hat{R}^{c_2d_2}_{a_2b_2}\cdots \hat{R}^{c_kd_k}_{a_kb_k}.
\ea
After considering the higher-curvature corrections, the solution in the theory will depend on the small parameter $\l$, i.e., $g_{ab}(\l)$, in which $\l=0$ describes the solution of Einstein gravity.

To evaluate $\math{F}(x, u, \l)$ on the null hypersurface $L$, we introduce the Gaussian null coordinate system $\{z, u, x\}$, in which the line element can be expressed as
\ba\begin{aligned}
ds^2(\l)=2(dz+z^2\a du+z \b_i dx^i)du+\g_{ij}dx^i dx^j\,,\quad\quad
\end{aligned}\ea
in which the null hypersurface $L$ is given by $z=0$, and $\a$, $\b_i$ and $\g_{ij}$ are the function of $u,z,x,\l$. Here the index $i, j, k, l$ denotes the coordinate of the cross-section $B$. The null generator of $L$ is given by $k^a=(\pd/\pd u)^a$. Using this line element, the nonvanishing component of the Christoffel symbol on $L$ can be further obtained,
\ba\begin{aligned}
&\G^k{}_{ij}=\hat{\G}^k{}_{ij}\,,\quad \G^j{}_{ui}=K^j_i\,,\quad \G^j{}_{zi}=\bar{K}^j_i\,,\quad \G^1{}_{uz}=\frac{1}{2}\b^i\,,\\
&\G^u{}_{ij}=-\bar{K}_{ij}\,,\,\, \G^u{}_{ui}=-\frac{1}{2}\b_i\,,\,\, \G^z{}_{ij}=-\bar{K}_{ij}\,,\,\, \G^z{}_{zi}=\frac{1}{2}\b_i\,,
\end{aligned}\ea
in which $\hat{\G}^k{}_{ij}$ is the Christoffel symbol of the induced metric $\g_{ij}$, and
\ba\begin{aligned}
K_{ij}=\frac{1}{2}\pd_u \g_{ij}\,, \quad\quad \bar{K}_{ij}=\frac{1}{2}\pd_z \g_{ij}
\end{aligned}\ea
are the extrinsic curvature associated with the null vectors $(\pd/\pd u)^a$ and $(\pd/\pd z)^a$ separately. Further calculation gives
\ba\begin{aligned}\label{expR}
&R_{ij}^{kl}=\hat{R}_{ij}^{kl}-4K_{[i}^{[k}\bar{K}_{j]}^{l]}\,,\quad R^{zj}_{ui}=-\pd_u K_i^j-K_{i}^kK^{j}_k\,,\\
&R_{ui}^{jk}=-2D^{[j}K_i^{k]}+K_i^{[j}\b^{k]}\,,\quad R^{zi}_{jk}=-2D_{[j}K^i_{k]}+K^i_{[j}\b_{k]}
\end{aligned}\ea
on the hypersurface $L$. Using the above results and considering the symmetry of the generalized Kronecker tensor, it is not difficult to get
\ba\begin{aligned}
E_{ab}k^a k^b=R_u^z+\l\sum^{k_\text{max}}_{k=2}\a_k [E^{(k)}]^z_u
\end{aligned}\ea
on the null hypersurface $L$. Considering the antisymmetry of the generalized Kronecker tensor and using Eq. \eq{expR}, it is not hard to get
\ba\begin{aligned}
[E^{(k)}]^z_u=&\frac{k}{2^{k-1}}R^{zj}_{ui}\d^{j i_2 j_2\cdots i_k j_k}_{i l_2 m_2\cdots l_k m_k}R_{i_2j_2}^{l_2m_2}\cdots R_{i_kj_k}^{l_km_k}\\
&+\frac{k(k-1)}{2^{k-1}}R^{l_1m_1}_{u j_1} R^{z m_2}_{i_2 j_2}\d^{j_1\, i_2\, j_2\, i_3\, j_3\,\cdots\, i_k\, j_k}_{l_1 m_1 m_2l_3 m_3\cdots l_k m_k}R_{i_3j_3}^{l_3m_3}\cdots R_{i_kj_k}^{l_km_k}
\end{aligned}\ea
for $k\geq 2$.

Then, using the result
\ba\begin{aligned}
\pd_u \hat{R}_{ab}^{cd}=K^{[c}_eR_{ab}^{d]e}-2D_{[a}D^{[c}K_{b]}^{d]}\,,
\end{aligned}\ea
and together with Eq. \eq{exptheta}, we can further obtain
\ba\begin{aligned}
&\pd_u\Q=\pd_u\q+\l\pd_u K_b^a\sum^{k_\text{max}}_{k=2}\frac{k \a_k}{2^{k-1}}\d_{ac_2d_2\cdots c_kd_k}^{b a_2b_2\cdots a_k b_k}\hat{R}^{c_2d_2}_{a_2b_2}\cdots \hat{R}^{c_kd_k}_{a_kb_k}\\
&+\l K_b^a\pd_u\hat{R}^{c_2d_2}_{a_2b_2}\sum^{k_\text{max}}_{k=2}\frac{k(k-1) \a_k}{2^{k-1}}\d_{ac_2d_2c_3d_3\cdots c_kd_k}^{b a_2b_2a_3b_3\cdots a_k b_k}\hat{R}^{c_3d_3}_{a_3b_3}\cdots \hat{R}^{c_kd_k}_{a_kb_k}.
\end{aligned}\ea
Combing the above results, we have
\ba\begin{aligned}\label{expF}
\math{F}&=E_u^z+\pd_u\Q\\
&=-K_{ab}K^{ab}+\l (\hat{H}^b_a-H^b_a)\pd_u K_b^a-\l K_b^cK^a_cH_a^b\\
&+\l(2 D^d K_a^e-K_a^d\b^e)(2D_bK_c^f-K^f_b\b_c)P_{abc}^{def}\\
&+\l K_a^d(K^{e}_{\tilde{e}}R_{bc}^{f\tilde{e}}-2D_{b}D^{e}K_{c}^{f})\hat{P}_{abc}^{def}\,,
\end{aligned}\ea
in which we have denoted
\ba\begin{aligned}\label{vdf}
H^b_a&=\sum^{k_\text{max}}_{k=2}\frac{k \a_k}{2^{k-1}}\hat{\d}_{ac_2d_2\cdots c_kd_k}^{b a_2b_2\cdots a_k b_k}R^{c_2d_2}_{a_2b_2}\cdots R^{c_kd_k}_{a_kb_k}\,,\\
\hat{H}^b_a&=\sum^{k_\text{max}}_{k=2}\frac{k \a_k}{2^{k-1}}\hat{\d}_{ac_2d_2\cdots c_kd_k}^{b a_2b_2\cdots a_k b_k}\hat{R}^{c_2d_2}_{a_2b_2}\cdots \hat{R}^{c_kd_k}_{a_kb_k}\,,\\
P_{abc}^{def}&=\sum^{k_\text{max}}_{k=2}\frac{k(k-1) \a_k}{2^{k-1}}\hat{\d}_{abcc_3d_3\cdots c_kd_k}^{def a_cb_d\cdots a_k b_k}R^{c_3d_3}_{a_3b_3}\cdots R^{c_kd_k}_{a_kb_k}\,,\\
\hat{P}_{abc}^{def}&=\sum^{k_\text{max}}_{k=2}\frac{k(k-1) \a_k}{2^{k-1}}\hat{\d}_{abcc_3d_3\cdots c_kd_k}^{def a_cb_d\cdots a_k b_k}\hat{R}^{c_3d_3}_{a_3b_3}\cdots \hat{R}^{c_kd_k}_{a_kb_k}\,,
\end{aligned}\ea
in which
\ba\begin{aligned}
\hat{\d}_{a_1\cdots a_i}^{b_1\cdots b_i}=i! \g_{[a_1}^{b_1}\cdots \g_{a_i]}^{b_i}
\end{aligned}\ea
is the $i$th-order generalized Kronecker tensor on the cross-section $B$.

In the following, we would like to judge the sign of $\math{F}(\l)$ when the coupling constant $\l$ is regarded as a small parameter. If we consider the solution $g_{ab}(\l)$ which is an analytic function of $\l$, then we can expand $\math{F}(\l)$ by $\l$,
\ba\begin{aligned}
\math{F}(\l)=\math{F}+\l \d\math{F}+\frac{\l^2}{2}\d^2\math{F}+\cdots \,,
\end{aligned}\ea
in which we have introduced the notation
\ba\begin{aligned}
\d^{i}\h(x)=\left.\frac{\pd^i \h(x,\l)}{\pd \l^i}\right|_{\l=0}
\end{aligned}\ea
to denote the $i$th-order variation of the quantity $\h(x,\l)$, and the symbol without $\l$ denotes its counterpart of $\l=0$. \\

\noindent{\bf Zeroth-order approximation:}

First, we consider the zeroth-order approximation of $\l$. From Eq. \eq{expF}, we can obtain
\ba
\math{F}(\l)=-K^{ab}K_{ab}+\math{O}(\l)\,.
\ea
Considering the fact that $K_{ab}$ is a spatial tensor on $B$, we have $K_{ab}K^{ab}\geq 0$ and therefore $\math{F}(\l)\leq 0$ under the zeroth-order approximation of $\l$, which implies that the covariant entropy bound is satisfied under the zeroth-order approximation. This result is straightforward because the theory with $\l=0$ is just the Einstein gravity. However, it is worthy noting that there exists a zeroth-order optimal condition with $K_{ab}=0$ (here we denote $K_{ab}=K_{ab}|_{\l=0}$) on the null hypersurface $L$ such that the zeroth-order term of $\math{F}(\l)$ vanishes and the sign of $\math{F}(\l)$ cannot be determined by the zeroth-order approximation of $\l$. In this case, we need to consider the higher-order approximation of $\l$. Thus we next consider the first-order approximation of $\l$ under the zeroth-order optimal condition $K_{ab}=0$.\\

\noindent{\bf First-order approximation:}

From Eq. \eq{expF}, the first-order variation of $\math{F}(\l)$ gives
\ba\begin{aligned}
\d\math{F}&=-2K_{b}^a\d K_a^b+(\hat{H}^b_a-H^b_a)\pd_u K_b^a-K_b^cK^a_cH_a^b\\
&+(2 D^d K_a^e-K_a^d\b^e)(2D_bK_c^f-K^f_b\b_c)P_{abc}^{def}\\
&+K_a^d(K^{e}_{\tilde{e}}R_{bc}^{f\tilde{e}}-2D_{b}D^{e}K_{c}^{f})\hat{P}_{abc}^{def}\,.
\end{aligned}\ea
Noting that the zeroth-order optimal condition $K_{ab}=0$ on $L$ also implies
\ba\label{optimal00}
\pd_u K^b_a=D_c K^b_a=K_a^b=0
\ea
on $L$, we have $\d \math{F}=0$ under the zeroth-order approximation of $\l$. Then, we have
\ba\begin{aligned}
\math{F}(\l)=\math{F}+\l \d \math{F}+\math{O}(\l^2)=\math{O}(\l^2)\,,
\end{aligned}\ea
which indicates that the generalized covariant entropy bound is satisfied under the first-order approximation of $\l$. Moreover, this also implies that the sign of $\math{F}(\l)$ should be determined by the higher-order terms of $\l$.

Moreover, from Eq. \eq{expR}, we can see that
\ba\begin{aligned}
R_{ij}^{kl}=\hat{R}_{ij}^{kl}
\end{aligned}\ea
on the null hypersurface under the zeroth-order optimal condition $K_i^j=0$, which implies that $H^a_b=\hat{H}^a_b$ and therefore we have $\d \math{F}=0$.
Then, we have
\ba\begin{aligned}
\math{F}(\l)=\math{F}+\l \d \math{F}+\math{O}(\l^2)=\math{O}(\l^2)\,,
\end{aligned}\ea
which indicates that the generalized covariant entropy bound is satisfied under the first-order approximation of $\l$. Moreover, this also implies that the sign of $\math{F}(\l)$ should be determined by the higher-order terms of $\l$. \\

\noindent{\bf Second-order approximation:}

Considering the zeroth-order optimal condition  $K^b_a=\pd_u K^b_a=D_c K^b_a=0$ on $L$, the second-order variation of $\math{F}(\l)$ can be further obtained
\ba\begin{aligned}\label{ddF}
\d^2 \math{F}&=-2\d K_{ab}\d K^{ab}+(\hat{H}^b_a-H^b_a)\pd_u \d K_b^a\,.
\end{aligned}\ea
Moreover, from Eq. \eq{expR}, we can see that
\ba\begin{aligned}
R_{ij}^{kl}=\hat{R}_{ij}^{kl}
\end{aligned}\ea
on the null hypersurface under the zeroth-order optimal condition, which implies that $H^a_b=\hat{H}^a_b$ on $L$ and the second-term of Eq. \eq{ddF} vanishes. Then, we have
\ba\begin{aligned}
\math{F}(\l)=-\l^2 \d K_{ab}\d K^{ab}+\math{O}(\l^3)\,.
\end{aligned}\ea
This result shows that $\math{F}(\l)\leq 0$ under the second-order approximation of $\l$. Similarly, there also exists a second-order optimal condition with $\d K_{ab}=0$ on the null hypersurface $L$ such that the higher-order approximation should be further considered to judge the sign of $\math{F}(\l)$. \\

\noindent{\bf Third-order approximation:}

Under the zeroth-order and second-order optimal conditions, it is not hard to see
\ba\begin{aligned}
K_a^b=\pd_u K_a^b=&D_c K_a^b=0\,,\quad \d K_a^b=\pd_u \d K_a^b=D_c \d K_a^b=0\,,\\
&H_a^b=\hat{H}_a^b\,,\quad\quad \d H_a^b=\d \hat{H}_a^b\,
\end{aligned}\ea
on the null hypersurface $L$. Then, we can further find that $\d^3\math{F}=0$, and therefore we have $\math{F}(\l)=\math{O}(\l^4)$, which implies that $\math{F}(\l)$ vanishes under the third-order approximation and we need further consideration for the higher-order approximation. \\

\noindent{\bf $n$th-order approximation}

In the following, we would like to prove $\math{F}(\l)\geq 0$ under the $n$th-order approximation when the $(n-1)$th-order approximation of $\math{F}(\l)$ vanishes based on the mathematical induction. By concluding the first three-order results, it is equivalent to proving the following proposition: under the $n$th-order approximation of $\l$, we have
\ba
\math{F}(\l)=\frac{\l^n}{n!}\d^n\math{F}\geq 0\,,
\ea
when the  $(n-1)$th-order approximation of $\math{F}(\l)$ vanishes, i.e., $\d^i \math{F}=0$ for $i\leq n-1$. Then, $n$th-order approximation of $\math{F}(\l)$ vanishes demands
\ba\begin{aligned}\label{optimal2n}
\d^i K_a^b=0\,,\quad\quad \text{for $i\leq [n/2]$}\,,
\end{aligned}\ea
on the null hypersurface $L$, where $[n/2]$ denotes the integer part of $n/2$.\\

\noindent\textsl{Proof.} Obviously, the proposition is satisfied for $k=0,1,2$.\\

\noindent {\bf (Case of $n=2m$.)} We assume that the proposition is satisfied when $n=2 m$ for $m\geq 0$, i.e., we have
\ba\begin{aligned}
\d^{2m}\math{F}\geq 0\,,
\end{aligned}\ea
when $\d^i \math{F}=0$ for $i\leq 2 m-1$, and $\math{F}(\l)$ vanishes under the $2m$th-order approximation (i.e., $\d^i \math{F}=0$ for $i\leq 2 m$) demands
\ba\begin{aligned}\label{optimal1}
\d^i K_a^b=0\,,\quad\quad \text{for $i\leq m$}
\end{aligned}\ea
on $L$, which implies that $K_a^b(\l)=0$ vanishes under the $m$th-order approximation of $\l$ and thus $H_a^b(\l)=\hat{H}_a^b(\l)$ under the $m$th-order approximation. Therefore, we have
\ba\begin{aligned}\label{KH}
&K_a^b(\l)=\l^{m+1} \tilde{K}_a^b\,,\quad\quad \hat{H}^b_a(\l)-H^b_a(\l)=\l^{m+1} \math{H}^b_a\,,
\end{aligned}\ea
in which
\ba\begin{aligned}
\tilde{K}_a^b&=\sum^{\inf}_{i=m+1}\frac{\l^{i-m-1}}{i!}\d^{i} K_a^b\,,\\
\math{H}_a^b&=\sum^{\inf}_{i=m+1}\frac{\l^{i-m-1}}{i!}(\d^{i} \hat{H}_a^b-\d^{i} H_a^b)\,.
\end{aligned}\ea
With a straightforward calculation, we have
\ba\begin{aligned}\label{FFF}
\math{F}(\l)&=-\l^{2m+2}\tilde{K}_{ab}\tilde{K}^{ab}+\l^{2m+3} \math{H}^b_a\pd_u \tilde{K}_b^a-\l^{2m+3} \tilde{K}_b^c\tilde{K}^a_cH_a^b\\
&+\l^{2m+3}(2 D^d\tilde{K}_a^e-\tilde{K}_a^d\b^e)(2D_b\tilde{K}_c^f-\tilde{K}^f_b\b_c)P_{abc}^{def}\\
&+\l^{2m+3} \tilde{K}_a^d(\tilde{K}^{e}_{\tilde{e}}R_{bc}^{f\tilde{e}}-2D_{b}D^{e}\tilde{K}_{c}^{f})\hat{P}_{abc}^{def}\,.
\end{aligned}\ea
This result shows that $\math{F}(\l)=0$ under the $(2m+1)$th-order approximation of $\l$ under the optimal condition \eq{optimal1}, this also means that $\d^{2m+1}\math{F}=0$ would not lead to additional conditions than the $2m$th-order case, which actually gives the saturation condition \eq{optimal2n} with $n=2m+1$. These show that the proposition with $n=2m+1$ is satisfied. \\

\noindent {\bf (Case of $n=2m+1$.)} Assume that the proposition is satisfied when $n=2 m+1$ for $m\geq 0$, and the $(2m+1)$th-order approximation of $\math{F}(\l)$ vanishes demanding
\ba\begin{aligned}\label{condition222}
\d^i K_a^b=0\,,\quad\quad \text{for $i\leq m$}\,,
\end{aligned}\ea
which also leads to the results in Eq. \eq{KH} and therefore $\math{F}(\l)$ is given by Eq. \eq{FFF}. That is to say, under the $(2m+2)$th-order approximation, we have
\ba\begin{aligned}
\math{F}(\l)=-\frac{\l^{2m+2}}{[(m+1)!]^2}\d^{m+1}K_{a}^b \d^{m+1}K^a_b+\math{O}(\l^{2m+3})\,,
\end{aligned}\ea
which implies that $\math{F}(\l)\leq 0$ $(\d^{2m+2}\math{F}\leq 0)$ under the $(2m+2)$th-order approximation. Together with the condition \eq{condition222}, the saturation of $\d^{2m+2}\math{F}$ demands
\ba\begin{aligned}
\d^i K_a^b=0\,,\quad\quad \text{for $i\leq m+1$}\,.
\end{aligned}\ea
This is actually the proposition with $n=2m+2$, i.e., we have completed the proof. $\Box$\\

The above result shows $\math{F}(\l)$ is always nonpositive and thus the generalized covariant entropy bound associated with the JM entropy is valid under any higher-order approximation of $\l$.

\section{Conclusion and discussion}\label{sec5}

In this paper, we consider the generalized covariant entropy bound for the theory in which the Einstein gravity is perturbed by the higher-order Lovelock terms and introduce a small parameter $\l$ to characterize these perturbations. After considering the linearized second law of black holes in Lanczos-Lovelock gravity, the entropy bound in this theory is naturally proposed by replacing the Bekenstein-Hawking entropy with the JM entropy. Then, we showed that the key point to examine the validity of covariant entropy bound is to judge the sign of the quantity $\math{F}=H_{ab}k^a k^b+k^a \grad_a \Q$, and the entropy bound is satisfied if $\math{F}\leq 0$.  After assuming two physical assumptions and that the metric $g_{ab}$ is an analytic function of $\l$, we illustrated that the dominant term of $\math{F}(\l)$ is always nonpositive based on the mathematical induction, i.e., the generalized covariant entropy bound is valid under any higher-order approximation of $\l$. This indicates that the entropy bound using the JM entropy is strictly satisfied under the perturbation level of the higher-order Lovelock terms. From the discussion in Sec. V D of Ref. \cite{Matsuda:2020yvl}, we can see that the above result also indicates the validity of the generalized second law under the higher-order approximation of $\l$ for the theory where the Einstein gravity is perturbed by the higher-order Lovelock terms, this is a different result from the linearized second law of Lanczos-Lovelock gravity.

From the calculations presented in this paper, it is not hard to check that if we replace the Bekenstein-Hawking with Wald entropy formula instead of JM entropy formula in the entropy bound, we cannot show the nonnegativity of $\math{F}$ only using the assumptions given by the paper. This implies that the covariant entropy bound might be used to select the black hole entropy of the gravitational theory. Moreover, it is worth noting that our result is only suitable for the case where the higher-order Lovelock terms are regarded as some small corrections to Einstein gravity, what about the non-perturbation cases? From the discussion in Sec. \ref{sec3}, the key point to examine the entropy bound is also to check the sign of $\math{F}$ given by Eq. \eq{expF}. However, due to the complexity of the expression, it is difficult for us to judge its sign directly only based on the setups and assumptions in our paper. One of our future work is going to consider these cases. Furthermore, it is also interesting to extend the discussion into the theory in which the Einstein gravity is perturbed by other higher-curvature terms.

\section*{Acknowledgement}
Jie Jiang is supported by the GuangDong Basic and Applied Basic Research Foundation with Grant no. 217200003 and the Talents Introduction Foundation of Beijing Normal University with Grant no. 310432102. Ming Zhang is supported by the National Natural Science
Foundation of China with Grant No. 12005080.


\begin{thebibliography}{100}
\bibitem{Bekenstein:1980jp}
J.~D.~Bekenstein, ``A Universal Upper Bound on the Entropy to Energy Ratio for Bounded Systems,''
Phys. Rev. D \textbf{23}, 287 (1981).
\bibitem{Bousso:1999xy}
R.~Bousso, ``A Covariant entropy conjecture,'' JHEP \textbf{07}, 004 (1999).
\bibitem{Fischler:1998st}
W.~Fischler and L.~Susskind, ``Holography and cosmology,'' arXiv:hep-th/9806039 [hep-th].
\bibitem{Gao:2004mc}
S.~Gao and J.~P.~S.~Lemos, ``The Covariant entropy bound in gravitational collapse,''
JHEP \textbf{04}, 017 (2004).
\bibitem{Gao:2005bk}
S.~Gao and J.~P.~S.~Lemos, ``Local conditions for the generalized covariant entropy bound,''
Phys. Rev. D \textbf{71}, 084010 (2005).
\bibitem{Gao:2008zy}
S.~Gao and X.~Wu, ``Proof of the entropy bound on dynamical horizons,''
JHEP \textbf{08}, 005 (2008).
\bibitem{Pesci:2007rp}
A.~Pesci, ``From Unruh temperature to generalized Bousso bound,'' Class. Quant. Grav. \textbf{24}, 6219-6226 (2007).
\bibitem{Pesci:2008yy}
A.~Pesci, ``On the statistical-mechanical meaning of Bousso bound,'' Class. Quant. Grav. \textbf{25}, 125005 (2008).
\bibitem{Flanagan:1999jp}
E.~E.~Flanagan, D.~Marolf and R.~M.~Wald, ``Proof of classical versions of the Bousso entropy bound and of the generalized second law,''
Phys. Rev. D \textbf{62}, 084035 (2000).
\bibitem{Bousso:2003kb}
R.~Bousso, E.~E.~Flanagan and D.~Marolf, ``Simple sufficient conditions for the generalized covariant entropy bound,''
Phys. Rev. D \textbf{68}, 064001 (2003).
\bibitem{Strominger:2003br}
A.~Strominger and D.~M.~Thompson, ``A Quantum Bousso bound,'' Phys. Rev. D \textbf{70}, 044007 (2004).
\bibitem{A13}
 B.~Zwiebach, ``Curvature Squared Terms and String Theories,'' Phys.\ Lett.\  {\bf 156B}, 315 (1985).
\bibitem{A14}
 D.~J.~Gross and E.~Witten, ``Superstring Modifications of Einstein's Equations,'' Nucl.\ Phys.\ B {\bf 277}, 1 (1986).
\bibitem{A15}
 A.~Sen, ``Black Hole Entropy Function, Attractors and Precision Counting of Microstates,'' Gen.\ Rel.\ Grav.\  {\bf 40}, 2249 (2008).
\bibitem{A16}
A.~Dabholkar and S.~Nampuri, ``Quantum black holes,'' Lect.\ Notes Phys.\  {\bf 851}, 165 (2012).
\bibitem{Lanczos}
 C.~Lanczos, ``A Remarkable property of the Riemann Christoffel tensor in four dimensions,'' Annals Math. {\bf39}, 842 (1938).
\bibitem{Lovelock}
 D. Lovelock, ``The Einstein tensor and its generalizations,'' J. Math. Phys. {\bf 12}, 498 (1971).
\bibitem{Lovelock2}
 B. Zwiebach, `` Curvature squared terms and string theories,'' Phys. Lett. {\bf B156}, 315 (1985).
\bibitem{Kovacs:2020ywu}
  A.~D.~Kovacs and H.~S.~Reall, ``Well-posed formulation of Lovelock and Horndeski theories,'' Phys.\ Rev.\ D {\bf 101}, 124003 (2020).
\bibitem{Matsuda:2020yvl}
T.~Matsuda and S.~Mukohyama, ``Covariant entropy bound beyond general relativity,'' Phys. Rev. D \textbf{103}, no.2, 024002 (2021).
\bibitem{A17}
R.~M.~Wald, ``Black hole entropy is the Noether charge,'' Phys.\ Rev. D\ {\bf 48}, R3427 (1993).
\bibitem{A18}
V.~Iyer and R.~M.~Wald, ``Some properties of Noether charge and a proposal for dynamical black hole,'' Phys.\ Rev.\ D {\bf 50}, 846 (1994).
\bibitem{Jacobson:1993xs}
T.~Jacobson and R.~C.~Myers, ``Black hole entropy and higher curvature interactions,''
Phys. Rev. Lett. \textbf{70}, 3684-3687 (1993).
\bibitem{A25}
 A.~Chatterjee and S.~Sarkar, ``Physical Process First Law and Increase of Horizon Entropy for Black Holes in Einstein Gauss-Bonnet Gravity,'' Phys. Rev. Lett. {\bf 108}, 091301 (2012).
\bibitem{A26}
 S.~Kolekar, T.~Padmanabhan, and S.~Sarkar, ``Entropy increase during physical processes for black holes in Lanczos-Lovelock gravity,'' Phys.\ Rev.\ D \ {\bf86}, 021501 (2012).
\bibitem{A27}
 S.~Sarkar and A.~C.~Wall, ``Generalized second law at linear order for actions that are functions of Lovelock densities,'' Phys. Rev. D {\bf 88}, 044017 (2013).
 \bibitem{A28}
J.~Jiang and M.~Zhang, ``Entropy increases at linear order in scalar-hairy Lovelock gravity,''
JHEP \textbf{04}, 148 (2020)
\end{thebibliography}
\end{document}